# Predicting Essential Components of Signal Transduction Networks: A Dynamic Model of Guard Cell Abscisic Acid Signaling

Song Li[1], Sarah M. Assmann[1], Réka Albert[2]*

1 Biology Department, Pennsylvania State University, University Park, Pennsylvania, United States of America, 2 Physics Department, Pennsylvania State University, University Park, Pennsylvania, United States of America

Plants lose water and take in carbon dioxide through microscopic stomatal pores, each of which is regulated by a surrounding pair of guard cells. During drought, the plant hormone abscisic acid (ABA) inhibits stomatal opening and promotes stomatal closure, thereby promoting water conservation. Dozens of cellular components have been identified to function in ABA regulation of guard cell volume and thus of stomatal aperture, but a dynamic description is still not available for this complex process. Here we synthesize experimental results into a consistent guard cell signal transduction network for ABA-induced stomatal closure, and develop a dynamic model of this process. Our model captures the regulation of more than 40 identified network components, and accords well with previous experimental results at both the pathway and whole-cell physiological level. By simulating gene disruptions and pharmacological interventions we find that the network is robust against a significant fraction of possible perturbations. Our analysis reveals the novel predictions that the disruption of membrane depolarizability, anion efflux, actin cytoskeleton reorganization, cytosolic pH increase, the phosphatidic acid pathway, or $K^+$ efflux through slowly activating $K^+$ channels at the plasma membrane lead to the strongest reduction in ABA responsiveness. Initial experimental analysis assessing ABA-induced stomatal closure in the presence of cytosolic pH clamp imposed by the weak acid butyrate is consistent with model prediction. Simulations of stomatal response as derived from our model provide an efficient tool for the identification of candidate manipulations that have the best chance of conferring increased drought stress tolerance and for the prioritization of future wet bench analyses. Our method can be readily applied to other biological signaling networks to identify key regulatory components in systems where quantitative information is limited.

## Introduction

One central challenge of systems biology is the distillation of systems level information into applications such as drug discovery in biomedicine or genetic modification of crops. In terms of applications it is important and practical that we identify the subset of key components and regulatory interactions whose perturbation or tuning leads to significant functional changes (e.g., changes in a crop's fitness under environmental stress or changes in the state of malfunctioning cells, thereby combating disease). Mathematical modeling can assist in this process by integrating the behavior of multiple components into a comprehensive model that goes beyond human intuition, and also by addressing questions that are not yet accessible to experimental analysis.

In recent years, theoretical and computational analysis of biochemical networks has been successfully applied to well-defined metabolic pathways, signal transduction, and gene regulatory networks [1–3]. In parallel, high-throughput experimental methods have enabled the construction of genome-scale maps of transcription factor–DNA and protein–protein interactions [4,5]. The former are quantitative, dynamic descriptions of experimentally well-studied cellular pathways with relatively few components, while the latter are static maps of potential interactions with no information about their timing or kinetics. Here we introduce a novel approach that stands in the middle ground of the above-mentioned methods by incorporating the synthesis and dynamic modeling of complex cellular networks that contain diverse, yet only qualitatively known regulatory interactions.

We develop a mathematical model of a highly complex cellular signaling network and explore the extent to which the network topology determines the dynamic behavior of the system. We choose to examine signal transduction in plant guard cells for two reasons. First, guard cells are central components in control of plant water balance, and better

Abbreviations: ABA, abscisic acid; PP2C, protein phosphatase 2C; Atrboh, NADPH oxidase; $Ca^{2+}_{c}$, cytosolic $Ca^{2+}$ increase; CaIM, $Ca^{2+}$ influx across the plasma membrane; CIS, $Ca^{2+}$ influx to the cytosol from intracellular stores; CPC, cumulative percentage of closure; GCR1, G protein–coupled receptor 1; GPA1, heterotrimeric G protein α subunit 1; KAP, $K^+$ efflux through rapidly activating $K^+$ channels (AP channels) at the plasma membrane; KOUT, $K^+$ efflux through slowly activating outwardly-rectifying $K^+$ channels at the plasma membrane; NO, nitric oxide; NOS, nitric oxide synthase; PA, phosphatidic acid; ROS, reactive oxygen species understanding of their regulation is important for the goal of engineering crops with improved drought tolerance. Second, abscisic acid (ABA) signal transduction in guard cells is one of the best characterized signaling systems in plants: more than 20 components, including signal transduction proteins, secondary metabolites, and ion channels, have been shown to participate in ABA-induced stomatal closure. ABA induces guard cell shrinkage and stomatal closure via two major secondary messengers, cytosolic $Ca^{2+}$ ($Ca^{2+}{}_c$) and cytosolic pH (pH$_c$). A number of signaling proteins and secondary messengers have been identified as regulators of $Ca^{2+}$ influx from outside the cell or $Ca^{2+}$ release from internal stores; the downstream components responding to $Ca^{2+}$ are certain vacuolar and plasma membrane $K^+$ permeable channels, and anion channels in the plasma membrane [6,7]. Increases in cytosolic pH promote the opening of anion efflux channels and enhance the opening of voltage-activated outward $K^+$ channels in the plasma membrane [8–10]. Stomatal closure is caused by osmotically driven cell volume changes induced by both $K^+$ and anion efflux through plasma membrane–localized channels. Despite the wealth of information that has been collected regarding ABA signal transduction, the majority of the regulatory relationships are known only qualitatively and are studied in relative isolation, without considering their possible feedback or crosstalk with other pathways. Therefore, in order to synthesize this rich knowledge, one needs to assemble the information on regulatory mechanisms involved in ABA-induced stomatal closure into a system-level regulatory network that is consistent with experimental observations. Clearly, it is difficult to assemble the network and predict the dynamics of this system from human intuition alone, and thus theoretical tools are needed.

* To whom correspondence should be addressed. E-mail: ralbert@phys.psu.edu

We synthesize the experimental information available about the components and processes involved in ABA-induced stomatal closure into a comprehensive network, and study the topology of paths between signal and response. To capture the dynamics of information flow in this network we express synergy between pathways as combinatorial rules for the regulation of each node, and formulate a dynamic model of ABA-induced closure. Both in silico and in initial experimental analysis, we study the resilience of the signaling network to disruptions. We systematically sample functional and dynamic perturbations in network components and uncover a rich dynamic repertoire ranging from ABA hypersensitivity to complete insensitivity. Our model is validated by its agreement with prior experimental results, and yields a variety of novel predictions that provide targets on which further experimental analysis should focus. To our knowledge, this is one of the most complex biological networks ever modeled in a dynamical fashion.

## Results

### Extraction and Organization of Data from the Literature

We focus on ABA induction of stomatal closure, rather than ABA inhibition of stomatal opening, because these two processes, although related, exhibit distinct mechanisms, and there is substantially more information on the former process than on the latter in the literature. Experimental information about the involvement of a specific component in ABA-induced stomatal closure can be partitioned into three categories. First, biochemical evidence provides information on enzymatic activity or protein–protein interactions. For example, the putative G protein–coupled receptor 1 (GCR1) can physically interact with the heterotrimeric G protein α component 1 (GPA1) as supported by split-ubiquitin and coimmunoprecipitation experiments [11]. Second, genetic evidence of differential responses to a stimulus in wild-type plants versus mutant plants implicates the product of the mutated gene in the signal transduction process. For example, the ethyl methanesulfonate–generated ost1 mutant is less sensitive to ABA; thus, one can infer that the OST1 protein is a part of the ABA signaling cascade [12]. Third, pharmacological experiments, in which a chemical is used either to mimic the elimination of a particular component, or to exogenously provide a certain component, can lead to similar inferences. For example, a nitric oxide (NO) scavenger inhibits ABA-induced closure, while a NO donor promotes stomatal closure; thus, NO is a part of the ABA network [13]. The last two types of inference do not give direct interactions but correspond to pathways and pathway regulation. The existing theoretical literature on signaling is focused on networks where the first category of information is known, along with the kinetics of each interaction. However, the availability of such detailed knowledge is very much the exception rather than the norm in the experimental literature. Here we propose a novel method of representing qualitative and incomplete experimental information and integrating it into a consistent signal transduction network.

First, we distill experimental conclusions into qualitative regulatory relationships between cellular components (signaling proteins, metabolites, ion channels) and processes. For example, the evidence regarding OST1 and NO is summarized as both OST1 and NO promoting ABA-induced stomatal closure. We distinguish between positive and negative regulation by using the verbs "promote" and "inhibit," represented graphically as "→" and "—|," respectively, and quantify the severity of the effect by the qualifier "partial." A partial promoter's (inhibitor's) loss has less severe effects than the loss of a promoter (inhibitor), most probably due to other regulatory effects on the target node. Using these relations, we construct a database that contains more than 140 entries and is derived from more than 50 literature citations on ABA regulation of stomatal closure (Table S1). A number of entries in the database correspond to a component-to-component relationship, such as "A promotes B," which is mostly obtained by pharmacological experiments (e.g., applying A causes B response). However, the majority of the entries belong to the two categories of indirect inference described above, and are of the type "C promotes the process (A promotes B)." This kind of information can be obtained from both genetic and pharmacological experiments (e.g., disrupting C causes less A-induced B response, or applying C and A simultaneously causes a stronger B response than applying A only). There are a few instances of documented independence of two cellular components, which we identify with the qualifier "no relationship." Most of the information is derived from the model species *Arabidopsis thaliana*, but data from other species, mostly *Vicia faba*, are also included where comparable information from *Arabidopsis thaliana* is lacking.

### Assembly of the ABA Signal Transduction Network

To synthesize all this information into a consistent network, we need to determine how the different pathways

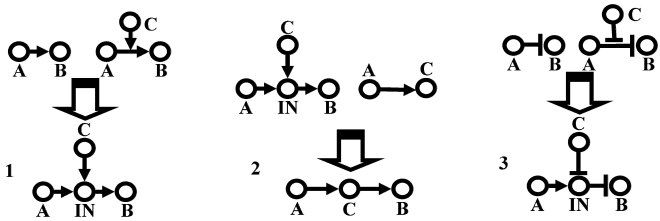

**Figure 1.** Illustration of the Inference Rules Used in Network Reconstruction

(1) If A → B and C → process (A → B), where A → B is not a biochemical reaction such as an enzyme catalyzed reaction or protein-protein/small molecule interaction, we assume that C is acting on an intermediary node (IN) of the A–B pathway.
(2) If A → B, A → C, and C → process (A → B), where A → B is not a direct interaction, the most parsimonious explanation is that C is a member of the A–B pathway, i.e. A → C → B.
(3) If A —| B and C —| process (A —| B), where A —| B is not a direct interaction, we assume that C is inhibiting an intermediary node (IN) of the A–B pathway. Note that A→ IN —| B is the only logically consistent representation of the A–B pathway.
DOI: 10.1371/journal.pbio.0040312.g001

suggested by experiments fit together (i.e., we need to find the pathways' branching and crossing points). We develop a set of rules compatible with intuitive inference, aiming to determine the sparsest graph consistent with all experimental observations. We summarize the most important rules in Figure 1; in the following we give examples for their application.

If A → B and C → process (A → B), where A → B is not a biochemical reaction such as an enzyme catalyzed reaction or protein–protein/small molecule interaction, we assume that C is acting on an intermediary node (IN) of the A–B pathway. This IN could be an intermediate protein complex, protein–small molecule complex, or multiple complexes (see Figure 1, panel 1). For example, ABA → closure, and NO synthase (NOS) → process (ABA → closure); therefore, ABA → IN → closure, NOS → IN. If A → B is a direct process such as a biochemical reaction or a protein–protein interaction, we assume that C → process (A → B) corresponds to C → A → B.

A → B and C → process (A → B) can be transformed to A → C → B if A → C is also documented. This means that the simplest explanation is to identify the putative intermediary node with C. For example, ABA → NOS, and NOS → process (ABA → NO) are experimentally verified and NOS is an enzyme producing NO, therefore, we infer ABA → NOS → NO (see Figure 1, panel 2).

A rule similar to rule 1 applies to inhibitory interactions (denoted by —|); however, in the case of A —| B, and C —| process (A —| B), the logically correct representation is: A → IN —| B, C —| IN (see Figure 1, panel 3).

The above rules constitute a heuristic algorithm for first expanding the network wherever the experimental relationships are known to be indirect, and second, minimizing the uncertainty of the network by filtering synonymous relationships. Mathematically, this algorithm is related to the problem of finding the minimum transitive reduction of a graph (i.e., for finding the sparsest subgraph with the same reachability relationships as the original) [14]; however, it differs from previously used algorithms by the fact that the edges can have one of two signs (activating and inhibitory), and edges corresponding to direct interactions are maintained.

In the reconstructed network, given in Figure 2, the network input is ABA and the output is the node "Closure." The small black filled circles represent putative intermediary nodes mediating indirect regulatory interactions. The edges (lines) of the network represent interactions and processes between two components (nodes); an arrowhead at the end of an edge represents activation, and a short segment at the end of an edge signifies inhibition. Edges that signify interactions derived from species other than *Arabidopsis* are colored light blue. We indicate two inferred negative feedback loops on S1P and $pH_c$ (see below) by dashed light blue lines. Nodes involved in the same metabolic reaction or protein complex are bordered by a gray box; only those arrows that point into or out of the box signify information flow (signal transduction). Some of the edges on Figure 2 are not explicitly incorporated in Table S1 because they represent general biochemical or physical knowledge (e.g., reactions inside gray boxes or depolarization caused by anion efflux).

A brief biological description of this reconstructed network (Figure 2) is as follows. ABA induces guard cell shrinkage and stomatal closure via two major secondary messengers, $Ca^{2+}_c$ and $pH_c$. Two mechanisms of $Ca^{2+}_c$ increase have been identified: $Ca^{2+}$ influx from outside the cell and $Ca^{2+}$ release from internal stores. $Ca^{2+}$ can be released from stores by InsP3 [15] and InsP6 [16], both of which are synthesized in response to ABA, or by cADPR and cGMP [17], whose upstream signaling molecule, NO [13,18], is indirectly activated by ABA. Opening of channels mediating $Ca^{2+}$ influx is mainly stimulated by reactive oxygen species (ROS) [19], and we reconstruct two ABA-ROS pathways involving OST1 [12] and GPA1 (L. Perfus-Barbeoch and S. M. Assmann, unpublished data), respectively. Based on current experimental evidence these two pathways are distinct, but not independent. The downstream components responding to $Ca^{2+}_c$ are certain vacuolar and plasma membrane $K^+$ permeable channels, and anion channels in the plasma membrane [6,7]. The mechanism of pH control by ABA is less clear, but it is known that $pH_c$ increases shortly after ABA treatment [20,21]. Increases in $pH_c$ levels promote the opening of anion efflux channels and enhance the opening of voltage-activated outward $K^+$ channels in the plasma membrane [8–10]. Stomatal closure is caused by osmotically driven cell volume changes induced by $K^+$ and anion efflux through plasma membrane-localized channels, and there is a complex interregulation between ion flux and membrane depolarization.

In addition to the secondary-messenger–induced pathways, there are two less-well-studied ABA signaling pathways involving the reorganization of the actin cytoskeleton, and the organic anion malate. ABA inactivates the small GTPase protein RAC1, which in turn blocks actin cytoskeleton disruption [22], contributing to an ABA-induced actin cytoskeleton reorganization process that is potentially $Ca^{2+}_c$ dependent [23]. In our model system, *Arabidopsis*, ABA regulation of malate levels has not been described. However, in *V. faba* it has been shown that ABA inhibits PEP carboxylase and malate synthesis [24], and that ABA induces malate breakdown [25]. In some conditions sucrose is an osmoticum that contributes to guard cell turgor [26,27] but no mechanisms of ABA regulation of sucrose levels have been described.

The recessive mutant of the protein phosphatase 2C (PP2C)

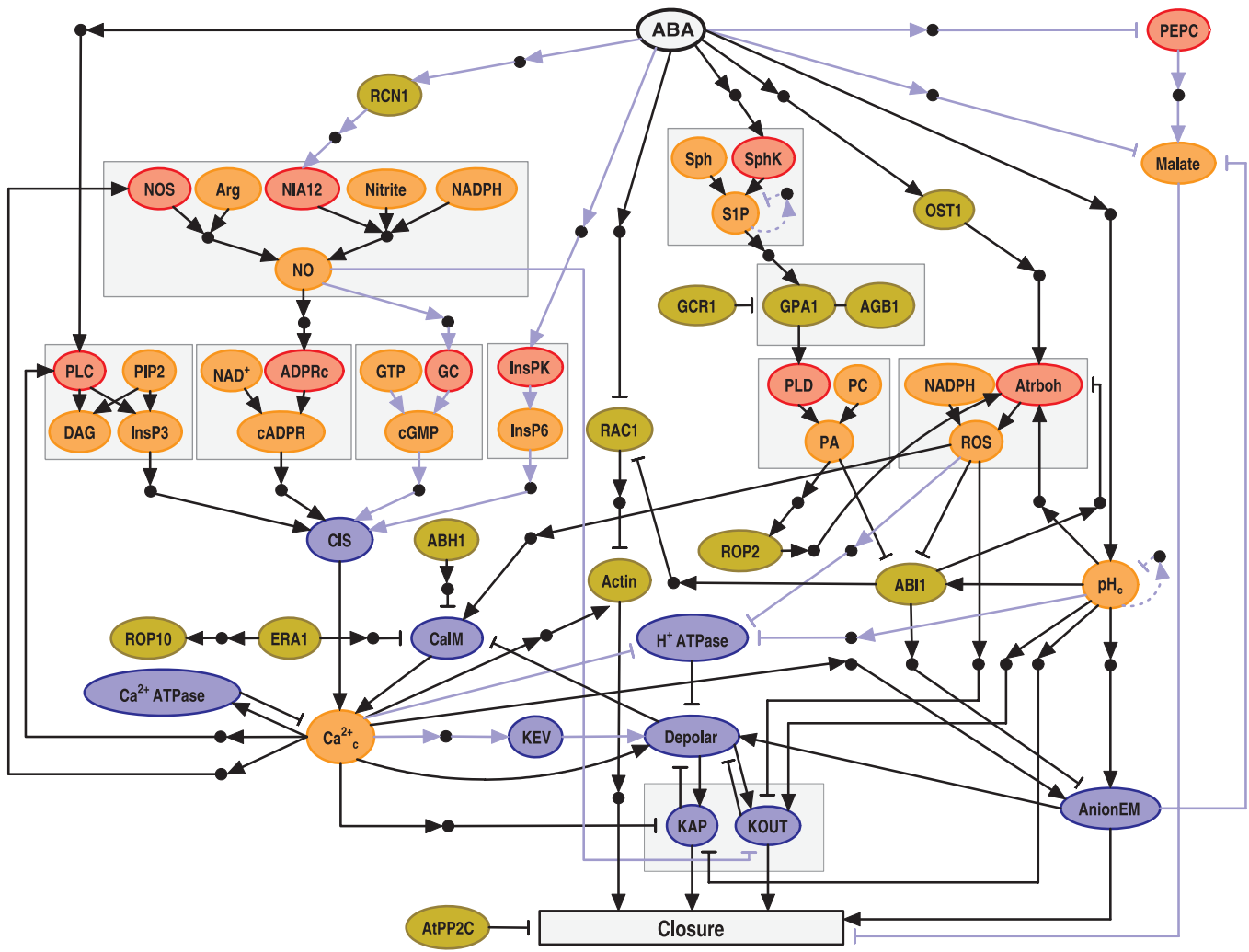

**Figure 2.** Current Knowledge of Guard Cell ABA Signaling

The color of the nodes represents their function: enzymes are shown in red, signal transduction proteins are green, membrane transport–related nodes are blue, and secondary messengers and small molecules are orange. Small black filled circles represent putative intermediary nodes mediating indirect regulatory interactions. Arrowheads represent activation, and short perpendicular bars indicate inhibition. Light blue lines denote interactions derived from species other than *Arabidopsis*; dashed light-blue lines denote inferred negative feedback loops on $pH_c$ and S1P. Nodes involved in the same metabolic pathway or protein complex are bordered by a gray box; only those arrows that point into or out of the box signify information flow (signal transduction).

The full names of network components corresponding to each abbreviated node label are: ABA, abscisic acid; ABI1/2, protein phosphatase 2C ABI1/2; ABH1, mRNA cap binding protein; Actin, actin cytoskeleton reorganization; ADPRc, ADP ribose cyclase; AGB1, heterotrimeric G protein β component; AnionEM, anion efflux at the plasma membrane; Arg, arginine; AtPP2C, protein phosphatase 2C; Atrboh, NADPH oxidase; CalM, $Ca^{2+}$ influx across the plasma membrane; $Ca^{2+}$ ATPase, $Ca^{2+}$ ATPases and $Ca^{2+}/H^+$ antiporters responsible for $Ca^{2+}$ efflux from the cytosol; $Ca^{2+}_c$, cytosolic $Ca^{2+}$ increase; cADPR, cyclic ADP-ribose; cGMP, cyclic GMP; CIS, $Ca^{2+}$ influx to the cytosol from intracellular stores; DAG, diacylglycerol; Depolar, plasma membrane depolarization; ERA1, farnesyl transferase ERA1; GC, guanyl cyclase; GCR1, putative G protein–coupled receptor; GPA1, heterotrimeric G protein α subunit; GTP, guanosine 5′-triphosphate; $H^+$ ATPase, $H^+$ ATPase at the plasma membrane; InsPK, inositol polyphosphate kinase; InsP3, inositol-1,4,5-trisphosphate; InsP6, inositol hexakisphosphate; KAP, $K^+$ efflux through rapidly activating $K^+$ channels (AP channels) at the plasma membrane; KEV, $K^+$ efflux from the vacuole to the cytosol; KOUT, $K^+$ efflux through slowly activating outwardly-rectifying $K^+$ channels at the plasma membrane; $NAD^+$, nicotinamide adenine dinucleotide; NADPH, nicotinamide adenine dinucleotide phosphate; NOS, Nitric oxide synthase; NIA12, Nitrate reductase; NO, Nitric oxide; OST1, protein kinase open stomata 1; PA, phosphatidic acid; PC, phosphatidyl choline; PEPC, phosphoenolpyruvate carboxylase; PIP2, phosphatidylinositol 4,5-bisphosphate; PLC, phospholipase C; PLD, phospholipase D; RAC1, small GTPase RAC1; RCN1, protein phosphatase 2A; ROP2, small GTPase ROP2; ROP10, small GTPase ROP10; ROS, reactive oxygen species; SphK, sphingosine kinase; S1P, sphingosine-1-phosphate.

DOI: 10.1371/journal.pbio.0040312.g002

ABI1, *abi1-1R*, is hypersensitive to ABA [28,29]. ABI1 is negatively regulated by phosphatidic acid (PA) and ROS, and $pH_c$ can activate ABI1 [30–32]. ABI1 negatively regulates RAC1 [22]. We hypothesize that ABI1 negatively regulates the NADPH oxidase (Atrboh) because ABI1 negatively regulates ROS production and Atrboh has been shown to be the dominant producer of ROS in guard cells [33]. We also assume that ABI1 inhibits anion efflux at the plasma membrane, because the dominant *abi1–1* mutant is known to affect the ABA response of anion channels [34] and because anion channels are documented key regulators of ABA-induced stomatal closure [35]. Components functioning downstream from ABI2 and its role in guard cell signaling are not well established, so ABI2 is not included. The newly isolated PP2C recessive mutants, *AtP2C-HA* [36] and *AtPP2CA* [37], exhibit minor ABA hypersensitivity. However, their downstream targets remain elusive; thus, we incorporate them as a general inhibitor of closure denoted AtPP2C.

Mutation of the gene encoding the mRNA cap-binding protein, ABH1, results in hypersensitivity of ABA-induced $Ca^{2+}_c$ elevation/oscillation and of anion efflux in plants grown under some environmental conditions [38,39]. We assume an inhibitory effect of ABH1 on $Ca^{2+}$ influx across the plasma membrane (CaIM), which can explain both of these effects due to the $Ca^{2+}$ regulation of anion efflux. Since the *abh1* mutation affects transcript levels of some genes involved in ABA response, this mutation may also affect ABA sensitivity by altering gene expression rather than by regulation of the rapid signaling events on which our network focuses. Mutations in the gene encoding the farnesyl transferase ERA1 or the gene encoding GCR1 also lead to hypersensitive ABA-induced closure; ERA1 has been shown to negatively regulate CaIM and anion efflux [40,41], whereas GCR1 has been shown to be interact with GPA1 [11]. We assume that ERA1 negatively regulates CaIM and GCR1 negatively regulates GPA1.

Another assumption in the network is that the protein phosphatase RCN1/PP2A regulates nitrate reductase (NIA12) activity as observed in spinach leaf tissue; this is expected to be a well-conserved mechanism due to the high sequence conservation of NIA-PP2A regulatory domains [42]. Figure 2 contains two putative autoregulatory negative feedback loops acting on S1P and $pH_c$, respectively. The existence of feedback regulation can be inferred from the published timecourse measurements of S1P [43] and $pH_c$ [21]—both indicating a fast increase in response to ABA, then a decrease—but the mediators are currently unknown. The assembled network is consistent with our biological knowledge with minimal additional assumptions, and it will serve as the starting point for the graph analysis and dynamic modeling described in the following sections.

## Modeling ABA Signal Transduction

Signaling networks can be represented as directed graphs where the orientation of the edges reflects the direction of information propagation (signal transduction). In a signal transduction network there exists a clear starting point, the node representing the signal (here, ABA), and one can follow the paths (successions of edges) that starting point to the node(s) representing the output(s) of the network (here, stomatal closure). The signal–output paths correspond to the propagation of reactions in chemical space, and can be thought of as pseudodynamics [44]. When only static information is available, pseudodynamics takes into account the graph theoretical properties of the signal transduction network. For example, one can measure the number of nodes or distinct network motifs that appear one, two,…n edges away from the signal node. Such motifs reflect different cellular signaling processing capabilities and provide important insights into the biological processes under investigation. Graph theoretical measures can also provide information about the importance (centrality) of signal mediators [45] and can predict the changes in path structure when nodes or edges in the network are disrupted. These disruptions, explored experimentally by genetic mutations, voltage-clamping, or pharmacological interventions, can be modeled in silico by removing the perturbed node and all its edges from the graph [46]. The absence of nodes and edges will disrupt the paths in the network, causing a possible increase in the length of the shortest path between signal (ABA) and output (closure), suggesting decreased ABA sensitivity, or in severe cases the loss of all paths connecting input and output (i.e., ABA insensitivity).

We find that there are several partially or completely independent (nonoverlapping) paths between ABA and closure. The path of pH-induced anion efflux is independent of the paths involving changes in $Ca^{2+}_c$. Based on the current knowledge incorporated in Figure 2, the path mediated by malate breakdown is independent of both $Ca^{2+}$ and pH signaling. This result could change if evidence of a suggested link between pH and malate regulation [47] is found; note that regulation of malate synthesis in guard cells appears to have cell-specific aspects [48]. Increase in $Ca^{2+}_c$ can be induced by several independent paths involving ROS, NO, or InsP6. Thanks to the existence of numerous redundant signal (ABA)–output (closure) paths, a complete disconnection of signal from output (loss of all the paths) is possible only if four nodes, corresponding to actin reorganization, $pH_c$ increase, malate breakdown, and membrane depolarization, are simultaneously disrupted. This indicates a remarkable topological resilience, and suggests that functionally redundant mechanisms can compensate for single gene disruptions and can maintain at least partial ABA sensitivity. However, path analysis alone cannot capture bidirectional signal propagation and synergy (cooperativity) in living biological systems. For example, two nonoverlapping paths that reach the node closure could be functionally synergistic. Using only path analysis, disruption of either path would not be predicted to lead to a disconnection of the signal (ABA) from the output (closure), but due to the synergy between the two paths, the closure response may be strongly impaired if either of the two paths is disrupted experimentally. Because of such limitations of path analysis, we turn from path analysis to a dynamic description.

Dynamic models have as input information (1) the interactions and regulatory relationships between components (i.e., the interaction network); (2) how the strength of the interactions depends on the state of the interacting components (i.e., the transfer functions); and (3) the initial state of each component in the system. Given these, the model will output the time evolution of the state of the system (e.g., the system's response to the presence or absence of a given signal). Given the incomplete characterization of the processes involved in ABA-induced stomatal closure (as is typical of the current state of knowledge of cell signaling cascades), we employ a qualitative modeling approach. We assume that the state of the network nodes can have two qualitative values: 0 (inactive/off) and 1 (active/on) [49]. These values can also describe two conformational states of a protein, such as closed and open states of an ion channel, or basal and high activity for enzymes. This assumption is necessary due to the absence of quantitative concentration or activity information for the vast majority of the network components. It is additionally justified by the fact that in the case of combinatorial regulation or cooperative binding, the input–output relationships are sigmoidal and thus can be distilled into two discrete output states [50].

Since "stomatal closure" does not usually entail the complete closure of the stomatal pore but rather a clear decrease in the stomatal aperture, and since there is a

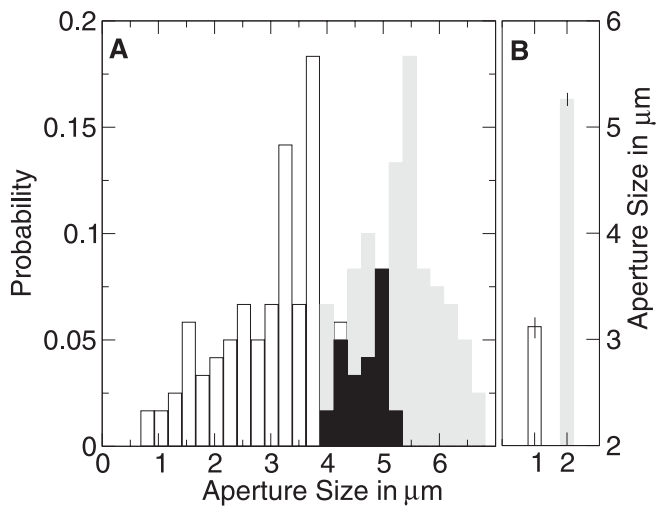

**Figure 3.** Stomatal Aperture Distributions without ABA Treatment (gray bars) and with 50 μM ABA (white bars)

(A) The x axis gives the stomatal aperture size and the y axis indicates the fraction of stomata for which that aperture size was observed. The black columns indicate the overlap between the 0 μM ABA and the 50 μM ABA distributions.

(B) Classical bar plot representation of stomatal aperture for treatment with 50 μM ABA (white bar, labeled 1) and without ABA treatment (gray bar, labeled 2) using mean ± standard error. This representation provides minimal information on population structure.

DOI: 10.1371/journal.pbio.0040313.g003

**Table 1.** Boolean Rules Governing the States of the Known (Named) Nodes in the Signal Transduction Network

| Node | Boolean Regulatory Rule |
|---|---|
| NO | NO* = NIA12 and NOS |
| PLC | PLC* = ABA and $Ca^{2+}_c$ |
| CaIM | CaIM* = (ROS or not ERA1 or not ABH1) and not Depolar |
| GPA1 | GPA1* = (S1P or not GCR1) and AGB1 |
| Atrboh | Atrboh* = $pH_c$ and OST1 and ROP2 and not ABI1 |
| $H^+$ ATPase | $H^+$ ATPase* = not ROS and not $pH_c$ and not $Ca^{2+}_c$ |
| Malate | Malate* = PEPC and not ABA and not AnionEM |
| RAC1 | RAC1* = not ABA and not ABI1 |
| Actin | Actin* = $Ca^{2+}_c$ or not RAC1 |
| ROS | ROS* = ABA and PA and $pH_c$ |
| ABI1 | ABI1* = $pH_c$ and not PA and not ROS |
| KAP | KAP* = (not $pH_c$ or not $Ca^{2+}_c$) and Depolar |
| $Ca^{2+}_c$ | $Ca^{2+}_c$* = (CaIM or CIS) and not $Ca^{2+}$ ATPase |
| CIS | CIS* = (cGMP and cADPR) or (InsP3 and InsP6) |
| AnionEM | AnionEM* = (($Ca^{2+}_c$ or $pH_c$) and not ABI1 ) or ($Ca^{2+}_c$ and $pH_c$) |
| KOUT | KOUT* = ($pH_c$ or not ROS or not NO) and Depolar |
| Depolar | Depolar* = KEV or AnionEM or not $H^+$ ATPase or not KOUT or $Ca^{2+}_c$ |
| Closure | Closure* = (KOUT or KAP ) and AnionEM and Actin and not Malate |

The nomenclature of the nodes is given in the caption of Figure 2. The nodes that have only one input are not listed to save space; a full description and justification can be found in Text S1. The next state of the node on the left-hand side of the equation (marked by *) is determined by the states of its effector nodes according to the function on the right-hand side of the equation.

DOI: 10.1371/journal.pbio.0040313.t001

significant variability in the response of individual stomata, the threshold separating the off (0) and on (1) state of the node "Closure" needs to invoke a population level description. We measured the stomatal aperture size distribution in the absence of ABA or after treatment with 50 μM ABA (see Materials and Methods). Our first observation was the population-level heterogeneity of stomatal apertures even in their resting condition (Figure 3A), a fact that may not be widely appreciated when more standard presentations, such as mean ± standard error, are used (see Figure 3B). The stomatal aperture distribution shifts towards smaller apertures after ABA treatment, and also broadens considerably. The latter result is inconsistent with the assumption that each stomate changing its aperture according to a common function that decreases with increasing ABA concentration, and suggests considerable cell-to-cell variation in the degree of response to ABA. Moreover, although there is a clear difference between the most probable "open" (0 ABA) and "closed" (+ ABA) aperture sizes, there also exists an overlap between the aperture size distribution of "open" and "closed" stomata. This result indicates the possibility of differential and cell-autonomous stomatal responses to ABA. In the absence of ± ABA measurements on the same stomate, we define the threshold of closure as a statistically significant shift of the stomatal aperture distribution towards smaller apertures in response to ABA signal transduction.

In our model the dynamics of state changes are governed by logical (Boolean) rules giving the state transition of each node given the state of its regulators (upstream nodes). We determine the Boolean transfer function for each node based on experimental evidence. The state of a node regulated by a single upstream component will follow the state of its regulator with a delay. If two or more pathways can independently lead to a node's activation, we combine them with a logical "or" function. If two pathways cannot work independently, we model their synergy as a logical "and" function. For nodes regulated by inhibitors we assume that the necessary condition of their activation (state 1) is that the inhibitor is inactive (state 0). As all putative intermediary nodes of Figure 2 are regulated by a single activator, and regulate a single downstream component, they only affect the time delays between known nodes; for this reason we do not explicitly incorporate intermediary nodes as components of the dynamic model. Table 1 lists the regulatory rules of known nodes of Figure 2; we give a detailed justification of each rule in Text S1.

Frequently in Boolean models time is quantized into regular intervals (timesteps), assuming that the duration of all activation and decay processes is comparable [51]. For generality we do not make this assumption, and in the absence of timing or duration information we follow an asynchronous method that allows for significant stochasticity in process durations [52,53]. Choosing as a timestep the longest duration required for a node to respond to a change in the state of its regulator(s) (also called a round of update, as each component's state will be updated during this time interval), the Boolean updating rules of an asynchronous algorithm can be written as:

$$S_i^n = B_i(S_j^{mj}, S_k^{mk}, S_l^{ml}, ..),  \qquad (1)$$

where $S_i^n$ is the state of component $i$ at timestep $n$, $B_i$ is the Boolean function associated with the node $i$ and its regulators $j,k,l,..$ and $mj, mk, ml, .. \in \{n - 1, n\}$, signifying that the timepoints corresponding to the last change in a input node's state can be in either the previous or current round of updates.

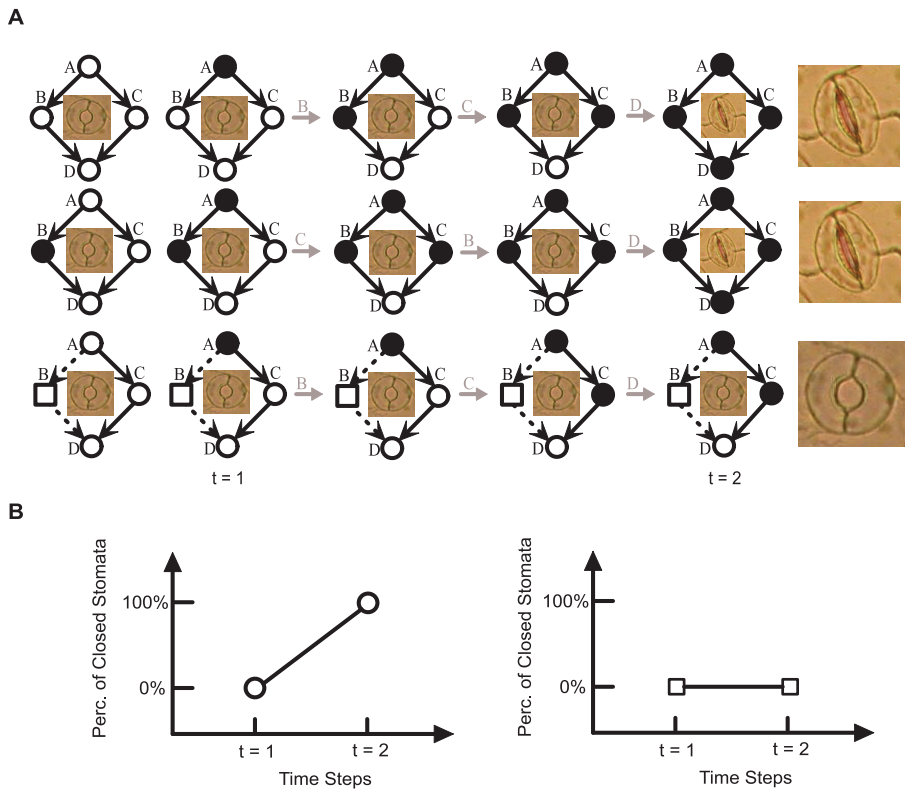

**Figure 4.** Schematic Illustration of Our Modeling Methodology and of the Probability of Closure

In this four-node network example, node A is the input (as ABA is the input of the ABA signal transduction network), and node D is the output (corresponding to the node "Closure" in the ABA signal transduction network). The nodes' states are indicated by the shading of their symbols: open symbols represent the off (0) state and filled symbols signify the on (1) state. To indicate the connection between this example and ABA-induced closure, we associate D = off (0) with a picture of an open stomate, and D = on (1) with a picture of a closed stomate. The Boolean transfer functions of this network are A* = 1, B* = A, C* = A, D* = B and C (i.e., node A is on commencing immediately after the initial condition, the next states of nodes B and C are determined by A, and D is on only when both B and C are on).
(A) The first column represents the networks' initial states; the input and output are not on, but some of the components in the network are randomly activated (e.g., middle row, node B). The input node A turns on right after initialization, signifying the initiation of the ABA signal. The next three columns in (A) represent the network's intermediary states during a sequential update of the nodes B, C, and D, where the updated node is given as a gray label above the gray arrow corresponding to the state transition. This sequence of three transitions represents a round of updates from timestep 1 (second column) to timestep 2 (last column). Out of a total of $2^2 \times 3! = 24$ possible different normal responses, two sketches of normal responses are shown in the top two rows. The bottom row illustrates a case in which one node (shown as a square) is disrupted (knocked out) and cannot be regulated or regulate downstream nodes (indicated as dashed edges).
(B) The probability of closure indicates the fraction of simulations where the output D = 1 is reached in each timestep; thus, in this illustration the probability of closure for the normal response (circles) increases from 0% at time step 1 to 100% at timestep 2. The knockout mutant's probability of closure (squares) is 0% at both time steps.
DOI: 10.1371/journal.pbio.0040312.g004

The relative timing of each process is chosen randomly and is changed after each update round such that we are sampling equally among all possibilities (see Materials and Methods). This approach reflects the lack of experimental data on relative reaction speeds. The internal states of signaling proteins and the concentrations of small molecules are not explicitly known for each stomate, and components such as $Ca^{2+}_c$ and cell membrane potential show various states even in a homogenous experimental setup [54,55]. Accordingly, we sample a large number (10,000) of randomly selected initial states for the nodes other than ABA and closure (closure is initially set to 0), and let the system evolve either with ABA always on (1) or ABA always off (0). We quantify the probability of closure (equivalent to the percentage of closed stomata in the population) by the formula

$$P(closure)^t = \sum_{j=1}^{N} S^t_{closure}(j) / N \qquad (2)$$

where $S^t_{closure}(j)$ is the state of the node "Closure" at time $t$ in the $j$th simulation and $N$ is the total number of simulations, in our case 10,000. We illustrate the main steps of our simulation method in Figure 4.

As shown in Figure 5, in eight steps, the system shows complete closure in response to ABA. In contrast, without ABA, although some initial states lead to closure at the beginning, within six steps the probability of closure approaches 0. Initial theoretical analysis of the attractors (stable behaviors) of this nonlinear dynamic system confirms that when given a constant ABA = 1 input, the majority of nodes will approach a steady state value within three to eight steps. This steady-state value does not depend on the initial conditions. For example, OST1, PLC, and InsPK stabilize in the on state, and PEPC settles into the off state within the first timestep when ABA is consistently on. The exception is a set of 12 nodes, including $Ca^{2+}_c$, $Ca^{2+}$ ATPase, NO, $K^+$ efflux from the vacuole to the cytosol, and $K^+$ efflux through rapidly

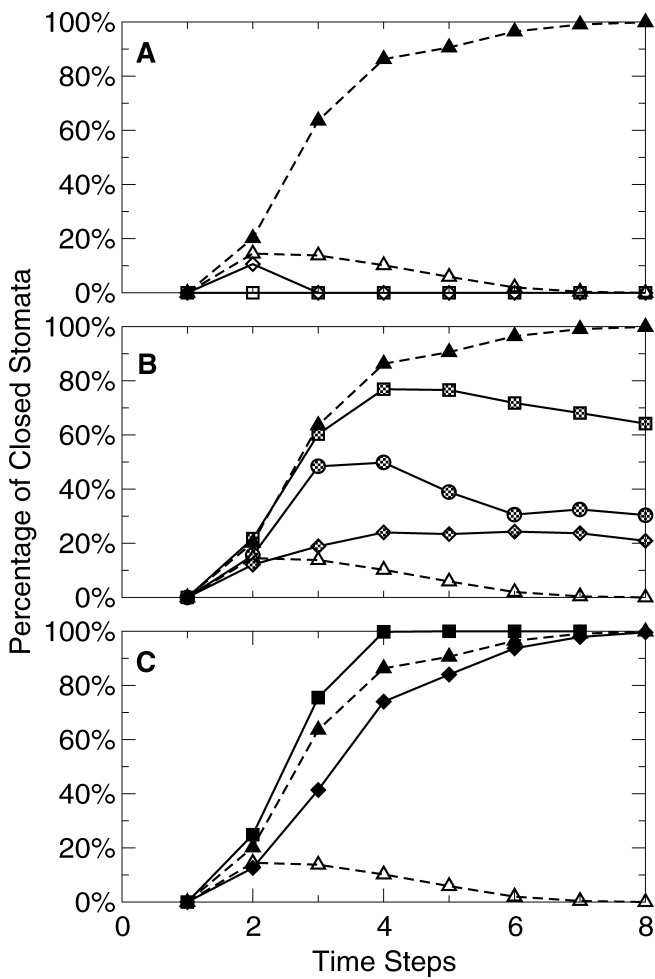

**Figure 5.** The Probability of ABA-Induced Closure (i.e., the Percentage of Simulations that Attain Closure) as a Function of Timesteps in the Dynamic Model

In all panels, black triangles with dashed lines represent the normal (wild-type) response to ABA stimulus. Open triangles with dashed lines show that in wild-type, the probability of closure decays in the absence of ABA.
(A) Perturbations in depolarization (open diamonds) or anion efflux at the plasma membrane (open squares) cause total loss of ABA-induced closure. The effect of disrupting actin reorganization (not shown) is identical to the effect of blocking anion efflux.
(B) Perturbations in S1P (dashed squares), PA (dashed circles), or pH$_c$ (dashed diamonds) lead to reduced closure probability. The effect of disrupting SphK is nearly identical to the effect of disrupting S1P (dashed squares); perturbations in GPA1 and PLD, KOUT are very close to perturbations in PA (dashed circles); for clarity, these curves are not shown in the plot.
(C) *abi1* recessive mutants (black squares) show faster than wild-type ABA-induced closure (ABA hypersensitivity). The effect of blocking Ca$^{2+}$ ATPase(s) (not shown) is very similar to the effect of the *abi1* mutation. Blocking Ca$^{2+}_c$ increase (black diamonds) causes slower than wild-type ABA-induced closure (ABA hyposensitivity). The effect of disrupting *atrboh* or ROS production (not shown) is very similar to the effect of blocking Ca$^{2+}_c$ increase.
DOI: 10.1371/journal.pbio.0040312.g005

activating K$^+$ channels (AP channels) at the plasma membrane (KAP), whose attractors are limit cycles (oscillations) according to the model. Ca$^{2+}_c$ oscillations have indeed been observed experimentally [56,57]; no time course measurements have been reported in the literature for the other components, so it is unknown whether they oscillate or not. We identified four subsets of behaviors for these nodes— distinguished by different positions on the limit cycle—depending on the initial conditions and relative process durations. Due to the functional redundancy between K$^+$ efflux mechanisms driving stomatal closure (see last entry of Table 1), and the stabilization of the other regulators of the node "Closure," a closed steady state (Closure = 1) is attained within eight steps for any initial condition. The details of this analysis will be published elsewhere.

### Identification of Essential Components

After testing the wild-type (intact) system, we investigate whether the disruption (loss) of a component changes the system's response to ABA. We systematically perturb the system by setting the state of a node to 0 (off state), and holding it at 0 for the duration of the simulation. This perturbation mimics the effect of a knockout mutation for a gene or pharmacological inhibition of secondary messenger production or of kinase or phosphatase activity. We characterize the effect of the node disruption by calculating the percentage (probability) of closure response to a constant ABA signal at each time step and comparing it with the percentage of closure in the wild-type system.

The perturbed system's responses can be classified into five categories with respect to the system's steady state and the time it takes to reach the steady state. We designate responses identical or very close to the wild-type response as having normal sensitivity; in these cases the probability of closure reaches 100% within eight timesteps. Disruptions that cause the percentage of closed stomata to decrease to zero after the first few steps are denoted as conferring ABA insensitivity (in accord with experimental nomenclature). We observe responses where the probability of closure (the percentage of stomata closed at any given timestep) settles at a nonzero value that is less than 100%; we classify these responses as having reduced sensitivity. Finally, in two classes of behavior the probability of closure ultimately reaches 100%, but with a different timing than the normal response. We refer to a response with ABA-induced closure that is slower than wild-type as hyposensitivity, while hypersensitivity corresponds to ABA-induced closure that is faster than wild-type. Therefore, the perturbed system's responses can be classified into five categories in the order of decreasing sensitivity defect: insensitivity to ABA, reduced sensitivity, hyposensitivity, normal sensitivity, and hypersensitivity.

We find that 25 single node disruptions (65%; compare with Table 2) do not lead to qualitative effects: 100% of the population responds to ABA with timecourses very close to the wild-type response. In contrast, the loss of membrane depolarizability, the disruption of anion efflux, and the loss of actin cytoskeleton reorganization present clear vulnerabilities: irrespective of initial conditions or of relative timing, all simulated stomata become insensitive to ABA (Figure 5A). Indeed, membrane depolarization is a necessary condition of K$^+$ efflux, which is a necessary condition of closure, as is actin cytoskeleton reorganization and anion efflux. The individual disruption of seven other components—PLD, PA, SphK, S1P, GPA1, K$^+$ efflux through slowly activating K$^+$ channels (KOUT), and pH$_c$ increase —reduces ABA sensitivity, as the percentage of closed stomata in the population decreases to 20%—80% (see Figure 5B). At least five components (S1P, SphK, PLD, PA, pH$_c$) of these 7 predicted components have been shown to impair ABA-

**Table 2.** Single to Triple Node Disruptions in the Dynamic Model

| Number of Nodes Disrupted | Percentage with Normal Sensitivity | Percentage Causing Insensitivity | Percentage Causing Reduced Sensitivity | Percentage Causing Hyposensitivity | Percentage Causing Hypersensitivity |
|---|---|---|---|---|---|
| 1 | 65% | 7.5% | 17.5% | 5% | 5% |
| 2 | 38% | 16% | 27% | 12% | 6% |
| 3 | 23% | 25% | 31% | 13% | 7% |

In all the perturbations, there are five groups of responses. Normal sensitivity refers to a response close to the wild-type response (shown as black triangles and dashed line in Figure 5). Insensitivity means that the probability of closure is zero after the first three steps (see Figure 5A). Reduced sensitivity means that the probability of closure is less than 100% (see dashed symbols in Figure 5B). Hyposensitivity corresponds to ABA-induced closure that is slower than wild-type (black diamonds in Figure 5C). Hypersensitivity corresponds to ABA-induced closure that is faster than wild-type (black squares in Figure 5C).
DOI: 10.1371/journal.pbio.0040312.t002

induced closure when clamped or mutated experimentally [8,31,43,58]. For these disruptions, both theoretical analysis and numerical results indicate that all simulated stomata converge to limit cycles (oscillations) driven by the $Ca^{2+}_c$ oscillations, yet the ratio of open and closed stomata in the population is the same at any timepoint, leading to a constant probability of closure. (The alternative possibility, of a subset of stomata being stably closed and another subset stably open, was not observed for any disruption.)

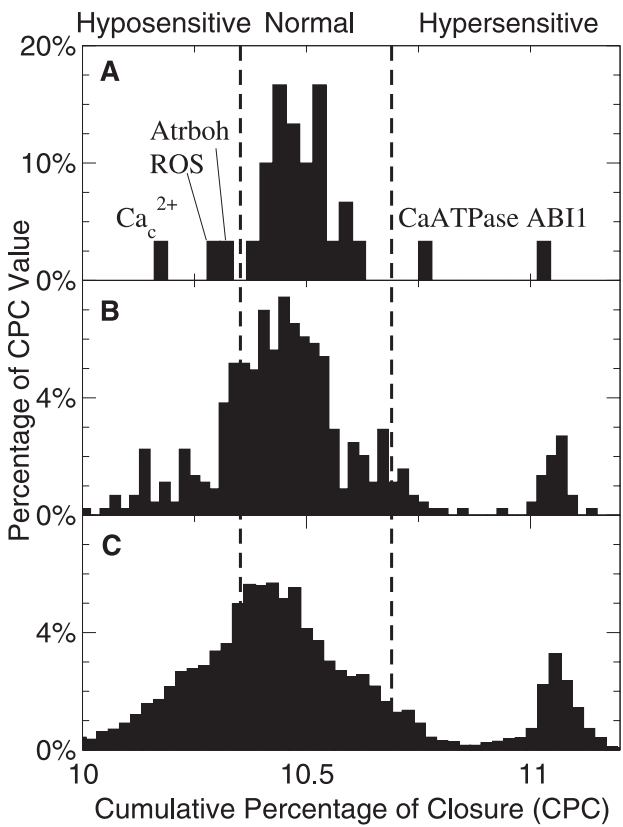

**Figure 6.** Classification of Close-to-Normal Responses

(A) For all the single mutants that ultimately reach 100% closure, we plot the histogram of the cumulative probability of closure (CPC). We find three distinct types of responses: hypersensitivity (CPC > 10.7, for *abi1* and $Ca^{2+}$ ATPase disruption); hyposensitivity (CPC < 10.35, for $Ca^{2+}_c$, *atrboh*, and ROS disruption); and normal responses ( 10.35 < CPC < 10.7). For all the double (B) and triple (C) mutants that eventually reach 100% closure at steady state when ABA = 1, we classify the responses using the CPC thresholds defined by the single mutant responses. The CPC threshold values are indicated by dashed vertical lines in the plot.
DOI: 10.1371/journal.pbio.0040312.g006

For all other single-node disruptions the probability of closure ultimately reaches 100% (i.e., all simulated stomata reach the closed steady state); however, the rate of convergence diverges from the rate of the wild-type response (see Figure 5C). Disruption of $Ca^{2+}_c$ increase or of the production of ROS leads to ABA hyposensitivity (slower than wild-type response). In contrast, the disruption of ABI1 or of the $Ca^{2+}$ ATPase(s) leads to ABA hypersensitivity (faster than wild type-response) (Figure 5C). The hyposensitive and hypersensitive responses are statistically distinguishable ($p < 0.05$ for all intermediary time steps [i.e., for $0 < t < 8$]) from the normal responses. Our model predicts that perturbation of OST1 leads to a slower than normal response that is nevertheless not slow enough to be classified as hyposensitive. Indeed, *ost1* mutants are still responsive to ABA even though not as strongly as wild-type plants [12].

After analyzing all single knockout simulations, we turned to analysis of double and triple knockout simulations. First, to effectively distinguish between normal, hypo- and hypersensitive responses (all of which achieve 100% probability of closure, but at different rates), we calculated the cumulative percentage of closure (CPC) by adding the probability of closure over 12 steps; the smaller the CPC value, the more slowly the probability of closure reaches 100%, and vice versa. Plotting the histogram of CPC values reveals a clear separation into three distinct groups of response in the case of single disruptions (Figure 6A). In contrast, the cumulative effects of multiple perturbations lead to a continuous distribution of sensitivities in a broad range around the normal (Figure 6B and 6C). We use the single perturbation results to identify three classes of response that achieve 100% closure, but at varying rates. We define two CPC thresholds: the midpoint between the most hyposensitive single mutant and normal response, $CPC_{hypo} = 10.35$; and the midpoint between the normal and least hypersensitive single mutant response, $CPC_{hyper} = 10.7$. Disruptions with cumulative closure probability < $CPC_{hypo}$ are classified as hyposensitive, disruptions with cumulative closure probability > $CPC_{hyper}$ are hypersensitive; and values between the two thresholds are classified as normal responses. This hypo/hypersensitive classification does not affect the determination of insensitive or reduced sensitivity responses, which are identified by observing a null or less than 100% probability of closure.

For double (triple) knockout simulations, some combinations of perturbations exhibit sensitivities that are independent of the sensitivity of each of their components' perturbation. Normal ABA-induced stomatal closure is

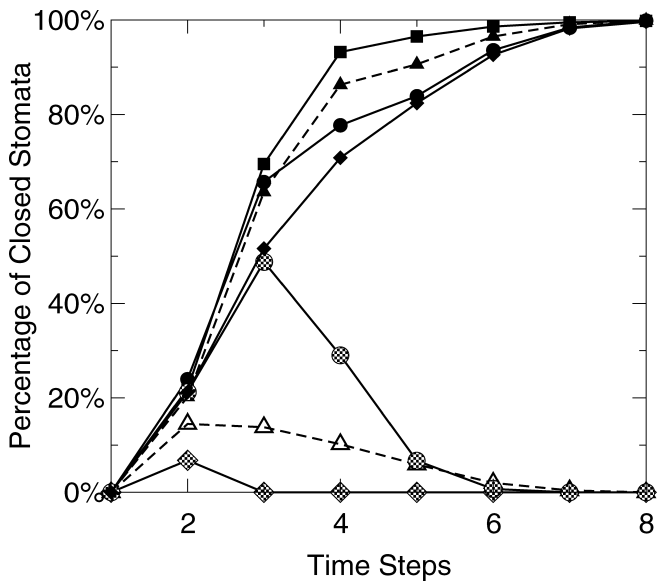

**Figure 7.** Summary of the Dynamic Effects of Calcium Disruptions

All curves represent the probability of ABA-induced closure (i.e., the percentage of simulations that attain closure) as a function of time steps. Black triangles with dashed line represent the normal (wild-type) response to ABA stimulus; open triangles with dashed lines show how the probability of closure decays in the absence of ABA. CIS + PA double mutants (dashed circles) and $Ca^{2+}_c$ + $pH_c$ double mutants (dashed diamonds) show insensitivity to ABA. $Ca^{2+}$ ATPase + RCN1 double mutants (black circles) show hyposensitive (delayed) response to ABA. Guanyl cyclase + CIS + CalM triple mutants (black diamonds) also show hyposensitivity; note that none of the guanyl cyclase or CIS or CalM single knockouts show changed sensitivity (data not shown). $Ca^{2+}$ ATPase mutants (black squares) show faster than wild-type ABA-induced closure (ABA hypersensitivity).

DOI: 10.1371/journal.pbio.0040312.g007

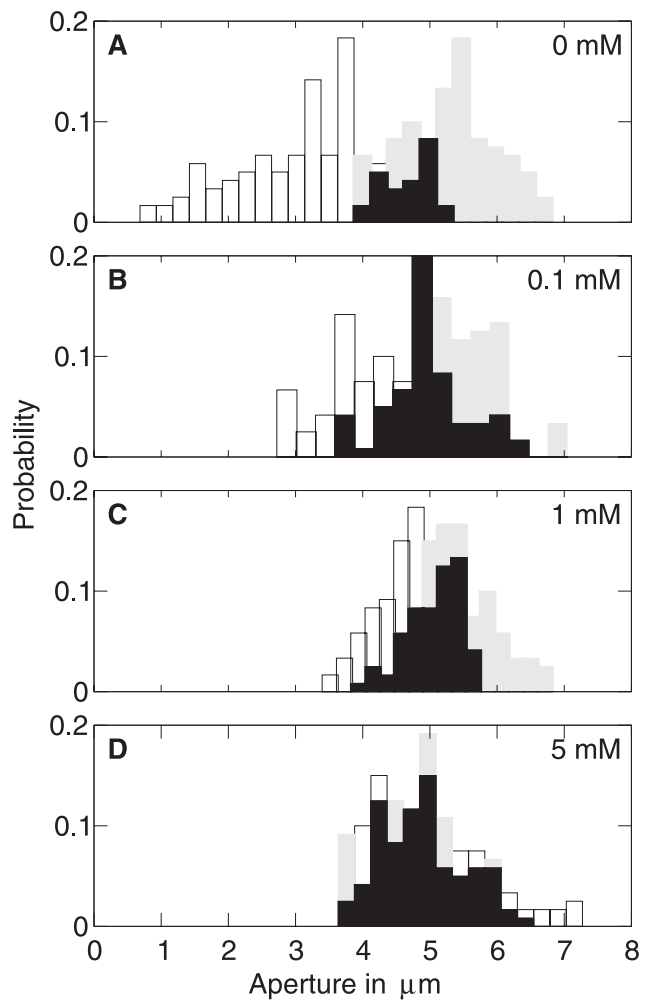

**Figure 8.** Effect of Cytosolic pH Clamp (Increasing Concentrations of Na-butyrate from 0 to 5 mM) on ABA-Induced Stomatal Closure

The histograms show the distribution of stomatal apertures without ABA treatment (gray bars) and with 50 μM ABA (white bars). Throughout, the x-axis gives the stomatal aperture size and the y-axis indicates the fraction of stomata for which that aperture size was observed. The black columns indicate the overlap between the 0 μM ABA and the 50 μM ABA distributions. Note that the data of (A) and those of Figure 3A are identical; these data are reproduced here for ease of comparison with panels (B–D).

DOI: 10.1371/journal.pbio.0040312.g008

preserved in 38% (23%) of combinations (see Table 2). In contrast, ABA signaling is completely blocked in 16% (25%) of disruptions. In addition to perturbations involving the three previously found insensitivity-causing single knockouts (loss of membrane depolarizability, the disruption of anion efflux, and the loss of actin cytoskeleton reorganization), a large number of novel combinations are found. Interestingly, perturbations of $Ca^{2+}$ or $Ca^{2+}$ release from stores, when combined with disruptions in PLD, PA, GPA1, or $pH_c$, lead to insensitivity (see Figure 7 and Discussion). ABA-induced closure is reduced (but not lost entirely) in 27% (31%) of the cases. Hyposensitive responses are found for 12% (13%) of double (triple) perturbations. All of the double perturbations in this category involve a knockout mutation of $Ca^{2+}_c$, Atrboh, or ROS. The triple perturbations involve a knockout mutation of $Ca^{2+}_c$, Atrboh, or ROS, plus two other perturbations, or combinations of three disruptions that alone are not predicted to cause quantifiable effects (e.g., guanyl cyclase, $Ca^{2+}$ release from internal stores [CIS], and CalM; see Figure 7). Around 6% (7%) of double (triple) perturbations, all including a knockout mutation of ABI1 or $Ca^{2+}$ ATPase, lead to a hypersensitive response. In summary, accumulating perturbations cause a dramatic decrease in the percentage of normal response; the majority of triple knockouts are either insensitive or have reduced sensitivity. The fraction of hyposensitive and hypersensitive knockouts increases only moderately.

## Experimental Assessment of Model Predictions

As a first step toward experimental assessment of the model's predictions, we used a weak acid, Na-butyrate, to clamp cytosolic pH, and then we treated the stomata with 50 μM ABA and observed the stomatal aperture responses. As shown in Figure 8A, the stomatal aperture distributions without butyrate treatments shift towards smaller apertures after ABA treatment, forming a distribution that overlaps with, but is clearly distinguishable from, the 0 ABA distribution. However, when increasing concentrations of butyrate are added in the solution, the "open" (0 ABA) and "closed" (+ ABA) distributions become increasingly overlapping (Figure 8B–8D). At the highest butyrate concentration (5 mM; Figure 8D), the 0 ABA and +ABA populations of stomatal apertures are statistically identical (the null hypothesis that the two distributions are the same cannot be rejected; two-tailed $t$ test, $p > 0.05$). These results qualitatively support our prediction of the importance of $pH_c$ signaling.

For a more quantitative comparison with the theoretically predicted probability of closure corresponding to pH clamping, one can define a threshold C between open and closed stomatal states, such that stomata with apertures larger than C can be classified as open and stomata with lower apertures can be classified as closed. We identify the threshold value C = 4.3 μm by simultaneously minimizing the fraction of stomata classified as closed in the control condition and maximizing this fraction in the ABA treated condition. Using this threshold we find that the fraction of closed stomata in the 50 μM ABA + 5 mM Na-butyrate population is 26%, in agreement with the theoretically predicted probability of closure (Figure 5B).

In plant systems, cytosolic pH changes in response to multiple hormones such as ABA [20,59], jasmonates [21], auxin [59], etc. The downstream effectors of pH changes include ion channels [8], protein kinases [60], and protein phosphatases [30]. Previous experiments with guard cells have demonstrated the efficacy of butyrate in imposing a cytosolic pH clamp [8,21]. While these prior experiments focused on a single concentration of butyrate, here we used five different concentrations (three shown), with 120 stomata sampled for each treatment. As seen in Figure 8, we were able to monitor the effect of butyrate in the +ABA treatment in both increasing the mean aperture size and reducing the spread of the aperture sizes. There is a clear indication of saturation between the two highest butyrate concentrations. While detailed measurements of cytosolic pH constitute a full separate study beyond the scope of the present article, the results of Figure 8 support the suggestion from our model that $pH_c$ should receive increased attention by experimentalists as a focal point for transduction of the ABA signal.

## Discussion

### Network Synthesis and Path Analysis

Logical organization of large-scale data sets is an important challenge in systems biology; our model provides such organization for one guard cell signaling system. As summarized in Table S1, we have organized and formalized the large amount of information that has been gathered on ABA induction of stomatal closure from individual experiments. This information has been used to reconstruct the ABA signaling network (Figure 2). Figure 2 uses different types of edges (lines) to depict activation and inhibition, and also uses different edge colors to indicate whether the information was derived from our model species, *Arabidopsis*, or from another plant species. Different types of nodes (metabolic enzymes, signaling proteins, transporters, and small molecules) are also color coded. An advantage of our method of network construction over other methods such as those used in Science's Signal Transduction Knowledge Environment (STKE) connection maps [61] is the inclusion of intermediate nodes when direct physical interactions between two components have not been demonstrated.

As is evident from Figure 2, network synthesis organizes complex information sets in a form such that the collective components and their relationships are readily accessible. From such analysis, new relationships are implied and new predictions can be made that would be difficult to derive from less formal analysis. For example, building the network allows one to "see" inferred edges that are not evident from the disparate literature reports. One example is the path from S1P to ABI1 through PLD. Separate literature reports indicate that PLDα null mutants show increased transpiration, that PLDα1 physically interacts with GPA1, that S1P promotion of stomatal closure is reduced in *gpa1* mutants, that PLD catalyses the production of PA, and that recessive *abi1* mutants are hypersensitive to ABA. Network inference allows one to represent all this information as the S1P → GPA1 → PLD → PA—| ABI1—| closure path, and make the prediction that ABA inhibition of ABI1 phosphatase activity will be impaired in sphingosine kinase mutants unable to produce S1P.

Another prediction that can be derived from our network analysis is a remarkable redundancy of ABA signaling, as there are eight paths that emanate from ABA in Figure 2 and, based on current knowledge (though see below) these paths are initially independent. The prediction of redundancy is consistent with previous, less formal analyses [62]. The integrated guard cell signal transduction network (which includes the ABA signal transduction network) has been proposed as an example of a robust scale-free network [62]. To classify a network as scale-free, one needs to determine the degree (the number of edges, representing interactions/regulatory relationships) of each node, and to calculate the distribution of node degrees (denoted degree distribution) [45,46]. Scale-free networks, characterized by a degree distribution described by a power law, retain their connectivity in the face of random node disruptions, but break down when the highest-degree nodes (the so-called hubs) are lost [46]. While the guard cell network may ultimately prove to be scale-free, the network is not sufficiently large at present to verify the existence of a power-law degree distribution; thus, the analogy with scale-free networks cannot be rigorously satisfied.

### Dynamic Modeling

Our model differs from previous models employed in the life sciences in the following fundamental aspects. First, we have reconstructed the signaling network from inferred indirect relationships and pathways as opposed to direct interactions; in graph theoretical terminology, we found the minimal network consistent with a set of reachability relationships. This network predicts the existence of numerous additional signal mediators (intermediary nodes), all of which could be targets of regulation. Second, the network obtained is significantly more complex than those usually modeled in a dynamic fashion. We bridge the incompleteness of regulatory knowledge and the absence of quantitative dose-response relationships for the vast majority of the interactions in the network by employing qualitative and stochastic dynamic modeling previously applied only in the context of gene regulatory networks [53].

Mathematical models of stomatal behavior in response to environmental change have been studied for decades [63,64]. However, no mathematical model has been formulated that integrates the multitude of recent experimental findings concerning the molecular signaling network of guard cells. Boolean modeling has been used to describe aspects of plant development such as specification of floral organs [65], and there are a handful of reports describing Boolean models of light and pathogen-, and light by carbon-regulated gene expression [66–68]. Use of a qualitative modeling framework for signaling networks is justified by the observation that signaling networks maintain their function even when faced with fluctuations in components and reaction rates [69]. Our model uses experimental evidence concerning the effects of gene knockouts and pharmacological interventions for inferring the downstream targets of the corresponding gene products and the sign of the regulatory effect on these targets. However, use of this information does not guarantee that the dynamic model will reproduce the dynamic outcome of the knockout or intervention. Indeed, all model ingredients (node states, transfer functions) refer to the node (component) level, and there is no explicit control over pathway-level effects. Moreover, the combinatorial transfer functions we employed are, to varying extents, conjectures, informed by the best available experimental information (see Text S1). Finally, in the absence of detailed knowledge of the timing of each process and of the baseline (resting) activity of each component, we deliberately sample timescales and initial conditions randomly. Thus, an agreement between experimental and theoretical results of node disruptions is not inherent, and would provide a validation of the model.

The accuracy of our model is indeed supported by its congruency with experimental observation at multiple levels. At the pathway level, our model captures, for example, the inhibition of ABA-induced ROS production in both *ost1* mutants and *atrboh* mutants [12,19,21] and the block of ABA-induced stomatal closure in a dominant-positive *atRAC1* mutant [22]. In our model, as in experiments, ABA-induced NO production is abolished in either *nos* single or *nia12* double mutants [13,18]. Moreover, the model reproduces the outcome that ABA can induce cytosolic $K^+$ decrease by $K^+$ efflux through the alternative potassium channel KAP, even when ABA-induced NO production leads to the inhibition of the outwardly-rectifying (KOUT) channel [70]. At the level of whole stomatal physiology, our model captures the findings that anion efflux [35,71] and actin cytoskeleton reorganization [22] are essential to ABA-induced stomatal closure. The importance of other components such as PA, PLD, S1P, GPA1, KOUT, $pH_c$ in stomatal closure control [8,20,31,43,58,72], and the ABA hypersensitivity conferred by elimination of signaling through ABI1 [28,29], are also reproduced. Our model is also consistent with the observation that transgenic plants with low PLC expression still display ABA sensitivity [73].

The fact that our model accords well with experimental results suggests that the inferences and assumptions made are correct overall, and enables us to use the model to make predictions about situations that have yet to be put to experimental test. For example, the model predicts that disruption of all $Ca^{2+}$ ATPases will cause increased ABA sensitivity, a phenomenon difficult to address experimentally due to the large family of calcium ATPases expressed in *Arabidopsis* guard cells (unpublished data). Most of the multiple perturbation results presented in Figure 5 and Table 2 also represent predictions, as very few of them have been tested experimentally. Results from our model can now be used by experimentalists to prioritize which of the multitude of possible double and triple knockout combinations should be studied first in wet bench experiments.

Most importantly, our model makes novel predictions concerning the relative importance of certain regulatory elements. We predict three essential components whose elimination completely blocks ABA-induced stomatal closure: membrane depolarization, anion efflux, and actin cytoskeleton reorganization. Seven components are predicted to dramatically affect the extent and stability of ABA-induced stomatal closure: $pH_c$ control, PLD, PA, SphK, S1P, G protein signaling (GPA1), and $K^+$ efflux. Five additional components, namely increase of cytosolic $Ca^{2+}$, Atrboh, ROS, the $Ca^{2+}$ ATPase(s), and ABI1, are predicted to affect the speed of ABA-induced stomatal closure. Note that a change in stomatal response rate may have significant repercussions, as some stimuli to which guard cells respond fluctuate on the order of seconds [74,75]. Thus our model predicts two qualitatively different realizations of a partial response to ABA: fluctuations in individual responses (leading to a reduced steady-state sensitivity at the population level), and delayed response. These predictions provide targets on which further experimental analysis should focus.

Six of the 13 key positive regulators, namely increase of cytosolic $Ca^{2+}$, depolarization, elevation of $pH_c$, ROS, anion efflux, and $K^+$ efflux through outwardly rectifying $K^+$ channels, can be considered as network hubs [45], as they are in the set of ten highest degree (most interactive) nodes. Other nodes whose disruption leads to reduced ABA sensitivity, namely SphK, S1P, GPA1, PLD, and PA, are part of the ABA → PA path. While they are not highly connected themselves, their disruption leads to upregulation of the inhibitor ABI1, thus decreasing the efficiency of ABA-induced stomatal closure. Similarly, the node representing actin reorganization has a low degree. Thus the intuitive prediction, suggested by studies in yeast gene knockouts [76,77], that there would be a consistent positive correlation between a node's degree and its dynamic importance, is not supported here, providing another example of how dynamic modeling can reveal insights difficult to achieve by less formal methods. This lack of correlation has also been found in the context of other complex networks [78].

Comparing Figure 3 and Figure 6C, one can notice a similar heterogeneity in the measured stomatal aperture size distributions and the theoretical distribution of the cumulative probability of closure in the case of multiple node disruptions. While apparently unconnected, there is a link between the two types of heterogeneity. Due to stochastic effects on gene and protein expression, it is possible that in a real environment not all components of the ABA signal transduction network are fully functional. Therefore, even genetically identical populations of guard cells may be heterogeneous at the regulatory and functional level, and may respond to ABA in slightly different ways. In this case, the heterogeneity in double and triple disruption simulations provides an explanation for the observed heterogeneity in the experimentally normal response: the latter is actually a mixture of responses from genetically highly similar but functionally nonidentical guard cells.

## Importance of $Ca^{2+}_c$ Oscillations to ABA-Induced Stomatal Closure

Through the inclusion of the nodes CaIM, CIS, and the $Ca^{2+}$ ATPase node representing the $Ca^{2+}$ ATPases and $Ca^{2+}/H^+$ antiporters [79,80] that drive $Ca^{2+}$ efflux from the cytosolic compartment, our model incorporates the phenom-enon of oscillations in cytosolic $Ca^{2+}$ concentration, which has been frequently observed in experimental studies [56,81,82]. In experiments where $Ca^{2+}_c$ is manipulated, imposed $Ca^{2+}_c$ oscillations with a long periodicity (e.g., 10 min of $Ca^{2+}$ elevation with a periodicity of once every 20 min) are effective in triggering and maintaining stomatal closure, yet at 10 min (i.e., after just one $Ca^{2+}_c$ transient elevation and thus before the periodicity of the $Ca^{2+}$ change can be "known" by the cell), significant stomatal closure has already occurred [56]. This result suggests that the $Ca^{2+}_c$ oscillation signature may be more important for the maintenance of closure than for the induction of closure [56,81], and that the induction of closure might only be dependent on the first, transient $Ca^{2+}_c$ elevation.

According to our model, if $Ca^{2+}_c$ elevation occurs, then stomatal closure is triggered (consistent with numerous experimental studies), but $Ca^{2+}_c$ elevation is not _required_ for ABA-induced stomatal closure. Re-evaluation of the experimental studies on ABA and $Ca^{2+}_c$ reveals support for this prediction. First, although $Ca^{2+}_c$ elevation certainly can be observed in guard cell responses to ABA, numerous experimental results also show that $Ca^{2+}_c$ elevation is only observed in a fraction of the guard cells assayed [9,83]. Furthermore, absence of $Ca^{2+}_c$ elevation in response to ABA does not prevent the occurrence of downstream events such as ion channel regulation [84,85] and stomatal closure [86,87], a phenomenon also predicted by our in silico analysis. Second, it has been observed that some guard cells exhibit spontaneous oscillations in $Ca^{2+}_c$, and in such cells, ABA application actually suppresses further $Ca^{2+}_c$ elevation [88]; thus, ABA and $Ca^{2+}_c$ elevation are clearly decoupled.

Our model does predict that disruption of $Ca^{2+}$ signaling leads to ABA hyposensitivity, or a slower than normal response to ABA. In the real-world environment, even a slight delay or change in responsiveness may have significant repercussions, as some stimuli to which guard cells respond fluctuate on the order of seconds; and stomatal responses can have comparable rapidity [74,75]. Moreover, our model predicts that $Ca^{2+}_c$ elevation (although not necessarily oscillation) becomes required for engendering stomatal closure when $pH_c$ changes, $K^+$ efflux or the S1P–PA pathway are perturbed (see Figure 7). Thus, $Ca^{2+}_c$ modulation confers an essential redundancy to the network. Support for such a redundant role can be found in a study by Webb et al. [89] where $Ca^{2+}$ concentration was reduced below normal resting levels by intracellular application of BAPTA (such reduction in baseline $Ca^{2+}_c$ levels has been shown to reduce ABA activation of anion channels [85]) and the epidermal tissue was perfused with $CO_2$-free air, a treatment that has been shown to inhibit outwardly rectifying $K^+$ channels and slow anion efflux channels [90]. The ABA insensitivity of stomatal closure found by Webb et al. under these conditions [89] therefore can be attributed to a combination of multiple perturbations (of $Ca^{2+}_c$ elevation, $K^+$ efflux, and anion efflux) and is consistent with the predictions of our model.

Our model indicates that double perturbations of the $Ca^{2+}$ ATPase component and either of RCN1, OST1, NO, NOS, NIA12, or Atrboh are hyposensitive (see Figure 7), consistent with experimental results on disruptions in the latter components [12,13,18,19,21,91]. Since the latter disruptions alone, with unperturbed $Ca^{2+}$ ATPase, are found to have a close-to-normal response in our model, a $Ca^{2+}$ ATPase–disrupted and therefore $Ca^{2+}_c$ oscillation–free model seems to be closer to experimental observations on stomatal aperture response recorded for these individual mutant genotypes. This suggests that $Ca^{2+}_c$ elevation (and not $Ca^{2+}_c$ oscillation) is the signal perceived by downstream factors that control the induction of closure. Possibly, certain as-yet-undiscovered interaction motifs, such as a synergistic feed-forward loop [92] or dual positive feedback loops [93], could transform the $Ca^{2+}_c$ oscillation into a stable downstream output.

## Limitations of the Current Analysis

**Network topology.** Our graph reconstruction is incomplete, as new signaling molecules will certainly be discovered. Novel nodes may give identity to the intermediary nodes that our model currently incorporates. Discovery of a new interaction among known nodes could simplify the graph by reducing (apparent) redundancy. For example, if it is found that GPA1 → OST1, the simplest interpretation of the ABA → ROS pathway becomes ABA → GPA1 → OST1 → ROS, and the graph loses one edge and an alternative pathway. As an effect, the graph's robustness will be attenuated. Among likely candidates for network reduction are the components currently situated immediately downstream of ABA because, in the absence of information about guard cell ABA receptors [94], we assumed that ABA independently regulates eight components. It is also possible that a newly found interaction will not change the existing edges, but only add a new edge. A newly added positive regulation edge will further increase the redundancy of signaling and correspondingly its robustness. Newly added inhibitory edges could possibly damage the network's robustness if they affect the main positive regulators of the network, especially anion channels and membrane depolarization. For example, experimental evidence indicates that _abi1 abi2_ double recessive mutants are more sensitive to ABA-induced stomatal closure than _abi1_ or _abi2_ single recessive mutants [29], suggesting that ABI1 and ABI2 act synergistically. Due to limited experimental evidence, we do not explicitly incorporate ABI2, but an independent inhibitory effect of ABI2 would diminish ABA signaling.

While it is difficult to estimate the changes in our conclusions due to future knowledge gain, we can gauge the robustness of our results by randomly deleting entries in Table S1 or rewiring edges of Figure 2 (see Texts S2 and S3). We find that most of the predicted important nodes are documented in more than one entry, and more than one entry needs to be removed from the database before the topology of the network related to that node changes (Text S2). Random rewiring of up to four edge pairs shows that the dynamics of our current network is moderately resilient to minor topology changes (Text S3 and Figure S1).

**Dynamic model.** In our dynamic model we do not place restrictions on the relative timing of individual interactions but sample all possible updates randomly. This approach reflects our lack of knowledge concerning the relative reaction speeds as well as possible environmental noise. The significance of our current results is the prediction that whatever the timing is, given the current topology of regulatory relationships in the network, the most essential regulators will not change. Our approach can be iteratively refined when experimental results on the strength and timing of individual interactions become available. For example, we can combine Boolean regulation with continuous synthesis and degradation of small molecules or signal transduction proteins [95,96] as kinetic (rate) data emerge. Our model considers the response of individual guard cell pairs to the local ABA signal; however, there is recent evidence of a synchronized oscillatory behavior of stomatal apertures over spatially extended patches in response to a decrease in humidity [97]. Our model can be extended to incorporate cell-to-cell signaling and spatial aspects by including extracellular regulators when information about them becomes available (see [51]).

**Node disruptions.** A knockout may either deprive the system of an essential signaling element (the gene itself), or it may "set" the entire system into a different state (e.g., by affecting the baseline expression of other, seemingly unrelated signaling elements). Our analysis and current experimental data only address the former. Because of this caveat, in some ways rapid pharmacological inhibition may actually have a more specific effect on the cell than gene knockouts.

## Implications

Many of the signaling proteins present as nodes in our model are represented by multigene families in Arabidopsis [98], with likely functional redundancy among encoded isoforms. Therefore, the amount of experimental work required to completely disrupt a given node may be considerable. It is also considerable work to make such genetic modification in many of the important crop species that are much less amenable than Arabidopsis to genetic manipulation. It is also the case that, at present, there are no reports of successful use of ratiometric pH indicators in the small guard cells of Arabidopsis, suggesting that further technical advances in this area are required. Facts such as these indicate the importance of establishing a prioritization of node disruption in experimental studies seeking to manipulate stomatal responses for either an increase in basic knowledge or an improvement in crop water use efficiency. Our model provides information on which such prioritization can be based. Future work on this model will focus on predicting the changes in ABA-induced closure upon constitutive activation of network components or in the face of fluctuating ABA signals. Ultimately, the experimental information obtained may or may not support the model predictions; the latter instance provides new information that can be used to improve the model. Through such iteration of in silico and wet bench approaches, a more complete understanding of complex signaling cascades can be obtained.

Approaches to describe the dynamics of biological networks include differential equations based on mass-action kinetics for the production and decay of all components [99,100], and stochastic models that address the deviations from population homogeneity by transforming reaction rates into probabilities and concentrations into numbers of molecules [101]. The great complexity of many cellular signal transduction networks makes it a daunting task to reconstruct all the reactions and regulatory interactions in such explicit biochemical and kinetic detail. Our work offers a roadmap for synthesizing incompletely described signal transduction and regulatory networks utilizing network theory and qualitative stochastic dynamic modeling. In addition to being the practical choice, qualitative dynamic descriptions are well suited for networks that need to function robustly despite changes in external and internal parameters. Indeed, several analyses found that the dynamics of network motifs crucial for the stable dynamics and noise-resistance of cellular networks, such as single input modules, feed-forward loops [102,103] and dual positive feedback loops [93], is correctly and completely captured by qualitative modeling [104,105]. For example, at the regulatory module level, several qualitative (Boolean and continuous/discrete hybrid) models [51,53,96] reproduced the *Drosophila* segment polarity gene network's resilience when facing variations in kinetic parameters [50], offering the most natural explanation of which parameter sets will succeed in forming the correct gene expression pattern [106]. We expect that our methods will find extensive applications in systems where modeling is currently not possible by traditional approaches and that they will act as a scaffold on which more quantitative analyses of guard cell signaling in particular and cell signaling in general can later be built.

Our analyses have clear implications for the design of future wet bench experiments investigating the signaling network of guard cells and for the translation of experimental results on model species such as Arabidopsis to the improvement of water use efficiency and drought tolerance in crop species [107–109]. Drought stress currently provides one of the greatest limitations to crop productivity worldwide [110,111], and this issue is of even more concern given current trends in global climate change [112,113]. Our methods also have implications in biomedical sciences. The use of systems modeling tools in designing new drugs that overcome the limitation of traditional medicine has been suggested in the recent literature [114]. Many human diseases, such as breast cancer [115] or acute myeloid leukemia [116,117], cause complex alterations to the underlying signal transduction networks. Pathway information relevant to human disease etiologies has been accumulated over decades and such information is stored in several databases such as TRANSPATH [118], BioCarta (http://www.biocarta.com), and STKE (http://www.stke.org). Our strategy can serve as a tool that guides experiments by integrating qualitative data, building systems models, and identifying potential drug targets.

## Materials and Methods

**Plant material and growth conditions.** Wild-type *Arabidopsis* (Col genotype) seeds were germinated on 0.5× MS media plates containing 1% sucrose. Seedlings were grown vertically under short-day conditions (8 h light/16 h dark) 120 μmol m$^{-2}$ s$^{-1}$ for 10 d. Vigorous seedlings were selected for transplantation into soil and were grown to 5 wk of age (from germination) under short day conditions (8 h light/16 h dark). Leaves were harvested 30 min after the lights were turned on in the growth chamber.

**Stomatal aperture measurements.** Leaves were incubated in 20 mM KCl, 5 mM Mes-KOH, and 1 mM CaCl$_2$ (pH 6.15) (Tris), at room temperature and kept in the light (250 μmol m$^{-2}$ s$^{-1}$) for 2 h to open stomata. For pH$_i$ clamping, different amounts of Na-butyrate stock solution (made up as 1M solution in water [pH 6.1]) were added into the incubation solution, to achieve the concentrations given in Figure 8, 15 min before adding 50 μM ABA. Apertures were recorded after 2.5 h of further incubation in light. Epidermal peels were prepared at the end of each treatment. The maximum width of each stomatal pore was measured under a microscope fitted with an ocular micrometer. Data were collected from 40 stomata for each treatment and each experiment was repeated three times.

**Model.** The network in Figure 2 was drawn with the SmartDraw software (http://www.smartdraw.com/exp/ste/home). The dynamic

modeling was implemented by custom Python code (http://www.python.org). To equally sample the space of all possible timescales, the random-order asynchronous updating method developed in [53] was used. Briefly, every node is updated exactly once during each unit time interval, according to a given order. This order is a permutation of the $N = 40$ nodes in the network, chosen randomly out of a uniform distribution over the set of all $N!$ possible permutations. A new update order is selected at each timestep. As demonstrated in [53], this algorithm is equivalent to a random timing of each node's state transition.

## Supporting Information

**Figure S1.** Probability of Closure in Randomized Networks where Pairs of Positive or Negative Edges Are Rewired

Found at DOI: 10.1371/journal.pbio.0040312.sg001 (677 KB EPS).

**Table S1.** Synthesis of Experimental Information about Regulatory Interactions between ABA Signal Transduction Pathway Components

Found at DOI: 10.1371/journal.pbio.0040312.st001 (407 KB DOC).

**Text S1.** Detailed Justification for Each Boolean Transfer Function

Found at DOI: 10.1371/journal.pbio.0040312.sd001 (149 KB DOC).

**Text S2.** Verification of the Inference Process and the Resulting Network

Found at DOI: 10.1371/journal.pbio.0040312.sd002 (45 KB DOC).

**Text S3.** Effect of Random Rewiring on the Network Dynamics

Found at DOI: 10.1371/journal.pbio.0040312.sd003 (36 KB DOC).

### Accession Numbers

The *Arabidopsis* Information Resource (TAIR) (http://www.arabidopsis.org) accession numbers for the genes discussed in this paper are *NIA12* (At1g77760/At1g37130), *GPA1* (At2g26300), *ERA1* (At5g40280), *AtrbohD/F* (At5g47910/At4g11230), *RCN1* (At1g25490), *OST1* (At4g33950), *ROP2* (At1g20090), *RAC1* (At4g35020), *ROP10* (At3g48040), *AtP2C-HA/AtPP2CA* (At1g72770/At3g11410), and *GCR1* (At1g48270).

## Acknowledgments

The authors thank Drs. Jayanth Banavar, Vincent Crespi, and Eric Harvill for critically reading a previous version of the manuscript; and Dr. István Albert for assistance with figure preparation.

**Author contributions.** SL, SMA, and RA conceived and designed the experiments. SL performed the experiments. SL and RA analyzed the data. SL, SMA, and RA wrote the paper.

**Funding.** RA gratefully acknowledges a Sloan Research Fellowship. Research on guard cell signaling in SMA's laboratory is supported by NSF-MCB02–09694 and NSF-MCB03–45251.

**Competing interests.** The authors have declared that no competing interests exist.